*Editorial*

# *Phase transitions* and ferroelectrics: revival and the future in the field

*Jens Kreisel* \*
Laboratoire Matériaux et Génie Physique
CNRS, Minatec, Grenoble Institute of Technology,
3 parvis Louis Néel, 38016 Grenoble, France

*Beatriz Noheda*
Zernike Institute for Advanced Materials
Department of Chemical Physics, University of Groningen
Nijenborgh 4, 9747 AG  Groningen, The Netherlands

*Brahim Dkhil*
Laboratoire Structures, Propriétés et Modélisation des Solides
Ecole Centrale Paris, CNRS-UMR 8580, 92290 Châtenay-Malabry, France

\* Jens.Kreisel@grenoble-inp.fr

## 1. Editorial - Introduction

As we celebrate *Phase Transitions*'s 30th anniversary this year, it is extremely rewarding to see tangible evidence of the Journal's continually increasing value to researchers in the field. We are pleased to announce that once again there has been an increase in the Journal's Impact Factor, to 1.201 for 2008 according to Thomson Reuters' Journal Citation Reports 2009. The increase of the impact factor, a measure of the average number of citations to a journal's articles, is a clear reflection of the dedicated support the journal receives from authors, editors and referees throughout the world; for this contribution, I would like to sincerely thank all the scientists involved.

Phase Transitions has been founded in 1979 by A.M. Glazer as the only journal devoted exclusively to this important subject, which remains one of the most challenging topics at the interface between condensed matter chemistry and condensed matter physics. Many of the most interesting physical phenomena are directly related to phase transitions in the solid state (superconductivity, ferromagnetism, ferroelectricity, multiferroicity, magnetoresistance, liquid crystals ...). Also most functional materials used in applications display specific properties that can be optimized/tuned, which is mainly due to the presence of phase transitions. The journal *Phase Transitions* provides a focus for papers on most aspects of phase transitions in condensed matter. Although emphasis is placed primarily on experimental work, theoretical papers are welcome if they have some bearing on experimental results. The areas of interest include:

- structural (ferroelectric, ferroelastic, high-pressure, order-disorder, Jahn-Teller, martensitic etc.) phase transitions
- multiferroics
- metal-insulator phase transitions
- magnetic phase transitions
- geophysical phase transitions
- superconducting and superfluid transitions
- critical phenomena and physical properties at phase transitions
- liquid crystals
- technological applications of phase transitions
- quantum phase transitions

By analyzing in some more detail the topics of papers published in *Phase Transitions* and the impact they have in the field, the editorial board of *Phase Transitions* has identified two particular fields of specific interest: functional oxides and liquid crystals. In the following we will discuss in some more detail the family of functional oxides while a subsequent editorial article will focus on liquid crystals.

Indeed, many of the papers published in Phase Transitions are related to functional oxides, i.e. oxides that posses a property/function which is - at least in principle – of interest for applications. Among functional oxides, we note a particular current interest for ferroelectrics and ferroelectric-related properties. The observed importance of ferroelectric materials for the journal *Phase Transitions* is in itself not surprising, since the class of *ferroic* materials, among which ferroelectrics, is by definition closely related to phase transitions [1] and exhibits an extraordinary range of structures, chemical bonding and physical properties. In account of this *Phase Transitions* has been happy to publish a number of focussed special issues in the field, whereof some directly deal with ferroelectric-related physics:

- Special issue on *Phase Transitions in Giant Piezoelectrics* [2]
- Special issue on *Current Progress in Multiferroics and Magnetoelectrics* [3]
- Special issue on *Ferroelectrics Physics* [4]

- Special Issue on *Structural and Ferroelectric Phase Transitions* 2006 [5] and 2008 [6]

, while others at least partly concern ferroelectrics:
- Special Issue o *Phase Transitions in Functional thin films* [7]
- Special issue on *Pressure-induced Phase Transitions* [8]
- Special Issue on *Computational Theory of Phase Transitions* [9]

After a first "Golden Age" of ferroelectrics, often related to the milestone book by Lines and Glass [10], we have seen in the past two decades new breakthroughs in the understanding of ferroelectrics, from the perspective of theory, experiments and applications. A recent book provides an excellent modern perspective of the latest developments in the physics of ferroelectrics [11]. We note that, although ferroelectrics have been known for a long time, new intriguing physical effects are indeed periodically discovered and their understanding remains an internationally highly-competitive area.

Having said all this, it appeared worthwhile us to present a state-of-the-art look at the field of ferroelectrics. We are certainly not attempting to provide a complete review of all aspects of the field of ferroelectrics over the last years but we wish to transport a flavour of the current excitement in the field through the (subjective) choice of four specific examples of current interest: (*i*) Piezoelectrics and the morphotropic phase boundary, (*ii*) Multiferroics, (*iii*) The effect of high pressure on ferroelectrics and (*iv*) Strain-engineering in ferroelectric oxide thin films. For each topic we will try to work out both current interesting approaches and an outlook into future challenges. Throughout our discussion, the reader is referred to a list of significant review articles, books and papers in the field, but the interested reader is encouraged to go beyond this starting point and we apologize to the authors whose relevant work is not cited in our restricted topical overview which does not aspire to provide a thorough review of the field.

## 2. Piezoelectrics, relaxors and the morphotropic phase boundary

Among the 32 crystallographic point groups, 20 can develop piezoelectricity. Piezoelectrics can also be pyroelectric (they show polarization) or ferroelectric (their polarization can be reversed under an electric field). Piezoelectrics are intrinsically multifunctional, leading to a wide range of applications ranging from medical diagnostics and therapy, industrial processing and process control, monitoring of environments, automotive and robotics, to multimedia or telecommunication. Piezoelectric materials are able to convert mechanical energy to electricity (and vice-versa): the larger the so-called electromechanical response, the more efficient is the application. Such ability is observed especially in ferroelectric solid solutions of perovskite-like structure, such as $PbZr_{1-x}Ti_xO_3$ (PZT) [12], which is the most intensively studied system. However, an even higher piezoelectric response is observed in relaxor-based single crystals like $Pb(Mg_{1/3}Nb_{2/3})O_3$-$PbTiO_3$ (PMN-PT) or $Pb(Zn_{1/3}Nb_{2/3})O_3$-$PbTiO_3$ (PZN-PT) [13]. Relaxors are characterized by a broad dielectric constant with a strong frequency dispersion of the temperature associated to the maximum of the dielectric constant, $T_{max}$. Intriguingly, contrary to normal ferroelectrics, $T_{max}$ is not connected to any structural phase transition. As a consequence, understanding the microscopic mechanisms responsible for both the giant piezoelectricity and the relaxor behaviour is an active and challenging area of research not only for applications but also for fundamental physics.

Large piezoelectric coefficients are observed in the vicinity of the so-called Morphotropic Phase Boundary (MPB), which was believed to separate rhombohedral and tetragonal phases. These phase do not have a crystallographic group-subgroup relationship and thus no continuous transition between them is possible. Today, our picture of the MPB has drastically changed. A real breakthrough was achieved experimentally [14-16] and theoretically (both phenomenological and using first-principles approaches) [17-19] by showing that the strong piezoelectric properties of these solid solutions is related to the "polarization rotation" between the adjacent rhombohedral and tetragonal phases through one (or more) intermediate phases of low-symmetry i.e. a monoclinic (orthorhombic or triclinic) phase. As a consequence, the previous phase diagrams have been revisited [20-24], showing the richness of the perovskite structure, which allows not only the rotations of the atomic displacement vectors but also changes in the oxygen rotation axes [25-27]. Interestingly, the polarization rotation can be achieved by an electric-field [28,29], pressure [30-32] or grain size modification [33-36] and it allows to design desired properties through the so-called domain-engineering. The discovery of the monoclinic MPB and the giant piezoelectric response in PMNT-PT have strongly contributed to the remarkable revival of the fundamental interest and activity on piezoelectric oxides (Figure 1).

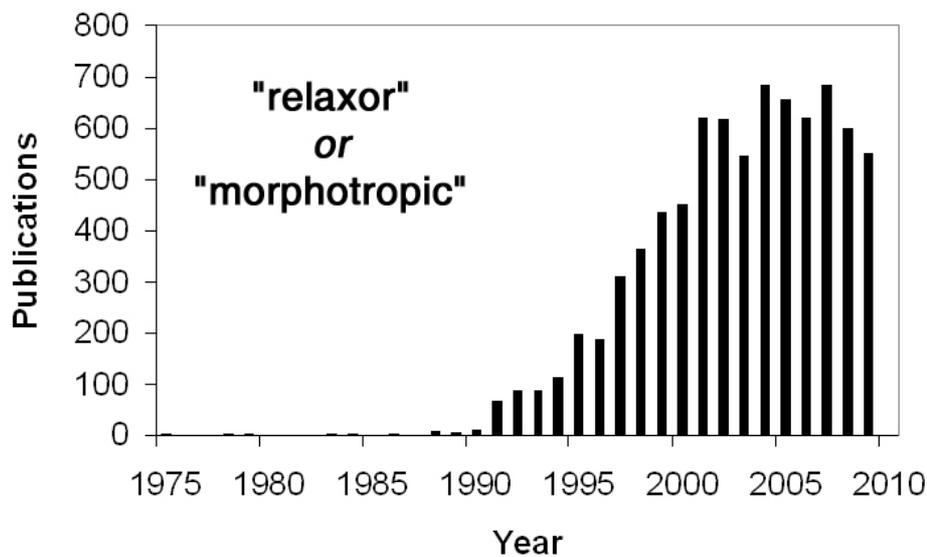

**Figure 1**
Evolution of the number of publications per year containing the keyword "relaxor" or "morphotropic" [37]

Another aspect for the understanding of giant piezoelectricity concerns the role played by the disorder. In complex oxides like PZT, PMN-PT etc. the disorder is of polar and/or chemical nature. It appears that the polar disorder of Pb cations plays a key role in these Pb-based solid solutions [38-41]. In case of relaxors, they are believed to form the so-called Polar Nano-Regions (PNRs), whose static and dynamic behaviour remains still unclear [42-52] and a matter of debate, especially with regard to the so-called waterfall effect [47,48]. As to the Pb-cations, let us emphasize that new environmental regulations of the European Union [53] provided more recently an added motivation for the research into Pb-free piezoelectrics, a field which has significantly increased from the year 2000 [54-63]. Nevertheless, despite a significant international effort in this direction, we are still far from the performances of PZT, PMN-PT etc., apart from some very few exceptions [59]. Future work into this field thus remains highly desirable.

Coming back to the disorder, not only lead but also the size and/or charge of the *B*-cations lead to chemical disorder in all above-mentioned solid solutions. Depending on how these cations are arranged, the structure and, thus, the properties are strongly affected [64,65]. This disorder is particularly important for relaxor-based materials and is believed to condition quenched random electric fields, which disturb the electric polarization and constitute a key ingredient for the understanding of relaxors [22,66-68].

More recently new ideas have emerged, introducing new concepts and questions which, in turn, ask for new experiments and modelling. The most significant new approaches are schematized in the following: *(i)* The MPB is considered rather as a region than as a boundary; *(ii)* The transformation pathway through the monoclinic phase and associated features such as the tweed-like domain patterns observed in MPB systems has lead some authors to consider that the monoclinic phase is an *adaptative phase* [69,70]. The adaptive phase is issued from martensitic-like phase transitions and is based on stress-accommodating twinned domains. According to this, the "monoclinic phase" observed on the average scale would not truly exist on the local scale but would be rather constructed by tetragonal and/or rhombohedral twinned nanodomains [71-76]. We note that some time ago, a similar idea was successfully used in the field of incommensurate and quasicrystal systems, leading to the debates about the real existence of (for instance) $5^{th}$-fold (pentagonal) symmetries or nano-twinning of $6^{th}$-fold twins. The existence of nanodomains at the MPB is expected both by the supporters of a true monoclinic phase and by the supporters of an adaptive phase. However, discerning the symmetry of the nanodomains has been an experimental challenge [77] *(iii)* It has been reported that the large piezoelectric coefficients are maximal not at zero field but rather at a finite field value associated to a critical end point, thus, to a *critical behaviour* similar to the liquid-gas critical point [78,79]. *(iv)* In the case of relaxor-based systems, it was proposed that the large piezoelectric response can be explained by a *proximity effect* [80] i.e. a coexistence of several close energy states at the nano-/-mesoscale similar to that proposed for colossal magneto-resistivity or high-temperature superconductivity [81,82]. These four recent concepts illustrate the richness of the current research in this field of giant piezoelectrics and the closeness with other systems displaying complex phase transitions. All these different ideas also highlight the need to bring together different scientific communities to advance in the field of phase transitions as such and complex piezoelectrics in particular.

Summarizing, a tremendous effort has been devoted to the study of oxides with a large piezoelectric response. Considerable progress has been achieved, as several breakthroughs and milestones have been accomplished. However, as highlighted above, a number of important challenges still remain unresolved. Among the most important aspects for future research are the tendency towards multiscale approaches and the research of new (preferably lead-free) piezoelectric compounds, both from the experimental, theoretical and the applications communities.

### 3. Multiferroics

The origin and understanding of coupling phenomena between different physical properties within one material is a central subject of solid state science. A great deal of theoretical and experimental attention in this field is currently focused on the coupling between magnetism and ferroelectricty as can be encountered in so-called multiferroics [83-93]. Multiferroics are multifunctional materials par excellence, because they posses simultaneously several so-called ferroic orders such as *ferro*magnetism, *ferro*electricity and/or *ferro*elasticity. The class of ferroics is also commonly extended to anti-ferroics: *antiferro*magnetism and *antiferro*electricity (an introduction and unified description of ferroic materials is given in ref. [1]). Ferroelectric-ferroelastic materials have been extensively studied and are at the origin of numerous applications with the showcase example PZT in which the coupling between deformation and electric polarization leads to a remarkable piezoelectric response (see section 2). Although ferroelectric-ferroelastic materials also belong to the class of multiferroics, the community restricts the term "multiferroic" currently rather to materials being both ferroelectric and magnetic, a class of materials which has been much less studied in the past due to their scarcity [83] but which currently experiences a truly remarkable increase in research effort (Figure 2).

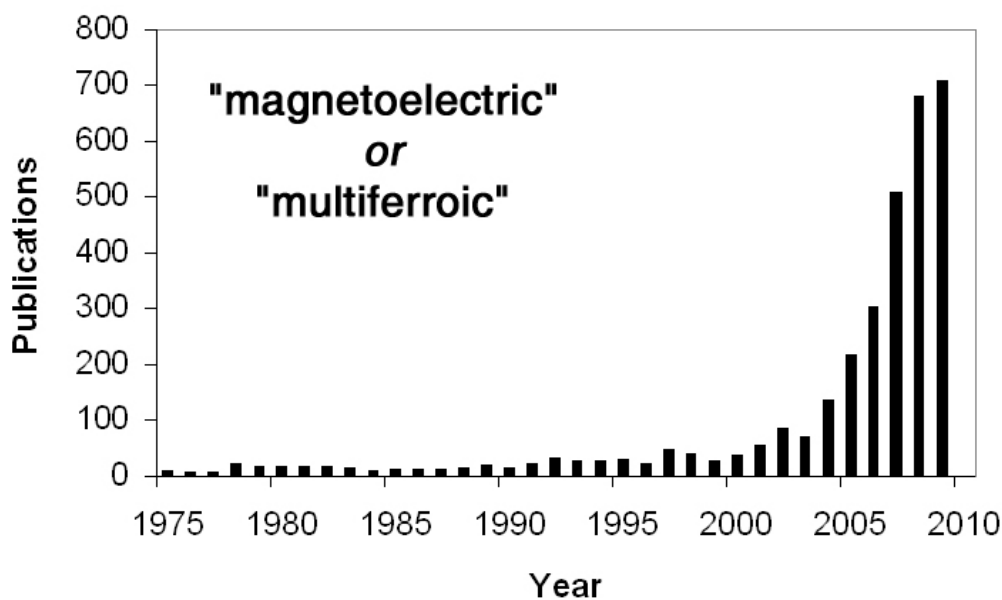

**Figure 2**
Evolution of the number of publications per year containing the keyword "magnetoelectric" or "multiferroic" [37]

Multiferroics can roughly be classified into three types of materials where ferroelectricity is conditioned by (i) hybridization effects, (ii) geometric constraints and/or (iii) electronic degrees of freedom (spin, charge or orbital). All three classes attract an intense research interest and present numerous challenges, as illustrated in the following:

(i) The multiferroic $BiFeO_3$ (BFO) is the prototype multiferroic with a ferroelectricity driven by hybridization effects and, even more importantly, it is perhaps the only material that is both magnetic and a ferroelectric with a strong electric polarization at 300K [94-98]. BFO has been synthesized in the form of thin films [95,99,100], ceramics [101] and crystals [102] some of which even up to centimeter-size [103]. Its high ferroelectric Curie ($T_C$ = 1083 K) and antiferromagnetic Néel ($T_N$ = 643K) temperatures together with its simple chemical formula, have indeed made BFO the preferred model system for fundamental and theoretical studies of multiferroics in the last few years. As a result, it has had an impact on the field of multiferroics that is comparable to that of yttrium barium copper oxide (YBCO) on superconductors [96]. Further to its fundamental interest, BFO is also considered to be one of the most promising candidates for realizing multiferroic spintronics devices [96,104-108]. A recent review [96] highlights in more detail the past and ongoing research efforts on BFO. Here we only note that, despite a considerable international effort, the model material BFO is still not completely understood, as exemplified by the fact that even the temperature-dependent [109-115] and pressure-dependent [96,110,112,116-119] phase sequences are still disputed in the literature. Future work into the P-T phase diagram beyond the 300K isotherm and the 1 bar isobar lines should help in getting a deeper understanding into the competing ferroelectric-ferroelastic-magnetic instabilities in BFO. A recent step forward in the understanding of BFO has been the observation of an electric-field-induced spin flop in BFO single crystals which demonstrates that the antiferromagnetic and ferroelectric order parameters in BFO are coupled

[120]. It has been argued that the magneto-electric coupling in BFO is intimately related to the presence of a magnetic cycloid structure that allows a coupling of ***M*** and ***P*** on an atomic level even in antiferromagnetic structures where on average the linear ME-effect is forbidden. We finally note that recent work on BFO highlights the interest of investigating fundamental excitations such as phonon-anomalies, spin-phonon-coupling, magnons, electromagnons etc. [112,121-129]. For instance, very recent Raman scattering measurements on $BiFeO_3$ single crystals show an important coupling between the magnetic order and lattice vibrations [122], which motivates a deeper look into such kind of coupling in other systems.

(ii) In geometric multiferroics, the ferroelectric instability is driven by size effects and/or other "structural" geometric considerations [130-132] and does not involve significant hybridization effects. The first material identified as a geometric ferroelectric was the hexagonal manganite $YMnO_3$ [130,132], a similar scheme has later be proposed [133] for Ba*M*$F_4$ fluorides and other systems are expected to be discovered. A major challenge remains to find such systems with a large electric polarization.

(iii) Rare-earth (*Re*) perovskite-type manganites are prototype multiferroics where ferroelectricity is driven by correlation effects and is strongly linked to electronic degrees of freedom such as spin-, charge-, or orbital-ordering [87,89,134-141]. Even though in such systems both ferroelectricty and magnetism have shown to be present, the ferroelectric polarisation in manganites is typically rather weak, mainly because the relativistic spin-orbit coupling is intrinsically weak [142]. In this class of multiferroics, magnetically-induced ferroelectrics such as $TbMnO_3$ or $Ni_3V_2O_8$ are probably the best studied and understood [140]. Here ferroelectricity arises from a complex magnetic spiral structure that breaks inversion symmetry and creates a polar axis. Magnetic order and ferroelectricity are thus directly coupled because there is no ferroelectric polarization without magnetic order.

Further to such systems *Re*-nickelates *Re*$NiO_3$ have more recently attracted interest and several authors [141-144] have proposed that coupling due to charge and magnetic order in the insulating regime might induce improper ferroelectricity. A very recent work [145] even predicts that *Re*-nickelates can exhibit a magnetically induced ferroelectric polarisation up to 10µC/cm², a polarisation two orders of magnitude larger than in multiferroic manganites. However, even though first confirming elements have been reported [144] a direct experimental evidence of multiferroicity in *Re*-nickelates is still missing and certainly deserves further attention due to its potential as magnetically-induced multiferroic system with a significant electric polarisation. In line with this , a ferroelectric polarization on the order of 1-10 µC/cm$^2$ has also been predicted for another class of magnetically-induced ferroelectrics which feature commensurate collinear magnetic order [146] : orthorhombic $HoMnO_3$ or $TmMnO_3$. Unfortunately these manganites, but also the nickelates are difficult or impossible to produce as single crystals making it difficult to test the theoretical predictions.

Up to now we have only discussed intrinsic multiferroics, i.e. where the different ferroic properties are united within *one* material (single-phase multiferroics), be it in the form of crystals, ceramics or thin films. To overcome the scarcity of single-phase multiferroics and to provide new magnetoelectric coupling mechanism recent work also concentrates on the class of *artificial* multiferroics in the form of composite-type materials or thin film nano-/hetero-structures [147-158]. Interestingly, such composite-type multiferroics, which incorporate both ferroelectric and magnetic phases, can yield giant magnetoelectric coupling response even above room temperature; an excellent review on this is given in [158]. In such systems, it is the elastic coupling interaction between the magnetostrictive phase and the piezoelectric phase that leads to the observed remarkable magnetoelectric response. Note that this mechanism is different to an intrinsic and direct coupling of ferroelectric and magnetic polarization as can be observed in single-phase magnetoelectric multiferroics. The field of composite-type multiferroic magnetoelectric is full of future possibilities and scientific challenges which deserve more attention.

Most of the research in multiferroics has been curiosity-driven basic research, but there are a number of device applications based on multiferroic materials. One of the more popular ideas is that multiferroic bits may be used to store information in the magnetization ***M*** and the polarization ***P***. The feasibility of such a 4 stage memory (two magnetic $M_{\uparrow\downarrow}$ and two ferroelectric $P_{\uparrow\downarrow}$) has been demonstrated recently [91]. Such a memory does not request the coupling between ferroelectricty and magnetism; a cross coupling would be even disastrous. If magneto-electric coupling is present, device applications could be realized where information is written magnetically, but stored in the electric polarization, leading to non-volatile memory. Multiferroics bits could also be used to increase the magnetic anisotropy to increase the decay time for magnetic storage. Other applications include magnetically field-tuned capacitors with which the frequency dependence of electronic circuits could be tuned with magnetic fields, or multiferroic sensors which measure magnetic fields through a zero-field current measurements.

The rapid development in the field of multiferroics in the last few years has been exciting and revealed fascinating physics in condensed matter. There are several important conclusions that we can draw regarding future directions of research. A first important question is what multiferroic mechanism leads to high ferroelectric polarization. Many of the investigated materials like the spin-spiral ferroelectrics are far away from charge instabilities, and magnetoelectric interactions are mediated by spin-orbit interactions. This combination naturally limits the ferroelectric polarization to relatively small values. Theory gives us a good guide in what type of materials we can expect larger polarizations and larger coupling effects. These are materials where magneto-electric effects are mediated by symmetric exchange or materials with intrinsic charge instabilities. However, experimental work has been limited due to a lack of novel materials.

A second important issue will be to develop further techniques allowing to provide experimental evidence for magnetoelectric coupling and to estimate its strength. One promising way may be the investigation of magnetic spin waves which couple to optical phonons (lattice vibrations) [122,125,127,159,160]. Such novel excitations - called electromagnons - are directly related to electromagnetic coupling and reflect the intimate relationship between magnetic and ferroelectric magnetic orders in multiferroic materials. The study of electromagnons may thus shed light on the strength of magnetoelectric coupling interactions. Electromagnons probably belong to the most challenging open questions in the field and are currently under intense investigation.

Another key question is in what kinds of other materials we can also expect multiferroic behavior. With a few exceptions, the multiferroics materials that have been investigated are transition metal oxides. It will be interesting to see how different types of magnetic and electric interactions found in rare-earth insulators without transition metal oxides, in organo-metallic materials, in magnetic chalcogenide-based phase-change material or in polymer-based magnets affect the magnetic and ferroelectric properties, and the magneto-electric coupling.

Finally, it remains a challenge to recreate some of the fascinating multiferroic properties in artificially engineered thin films and heterostructures. Without doubt, the study of such materials will be an important pillar of multiferroic research towards application and for basic research, particularly for key model materials that cannot be grown as clean single-crystals.

## 4. Effect of high pressure on ferroelectrics

In the past, much progress in understanding ferroelectrics has been achieved through temperature-, electric field- and/or chemical composition-dependent investigations. The use of pressure for the investigation of ferroelectrics has been relatively rare and was mostly limited to a pressure below 10 GPa [161-167] or a uniaxial pressure [168-173] due to the experimental difficulties, which have persisted in the past. The experimental difficulty in achieving high- (up to 30 GPa) and very high-pressure (say 100 GPa) has now been overcome for a number of years, which in turn led to an evolution of high-pressure studies from a "niche" activity reserved to some experts to an increasingly important activity (Figure 3). One of the attracting properties of the external parameter high-pressure is its character of a "cleaner" variable when compared to other parameters like temperature, since it acts only on interatomic distances [165]. Furthermore, an external pressure can notably modify the subtle energetical order between different phases in perovskite materials. Finally, experimental high-pressure data provide a serious test for structural ab-initio models and the implementation of calculations under pressure is much easier when compared to temperature.

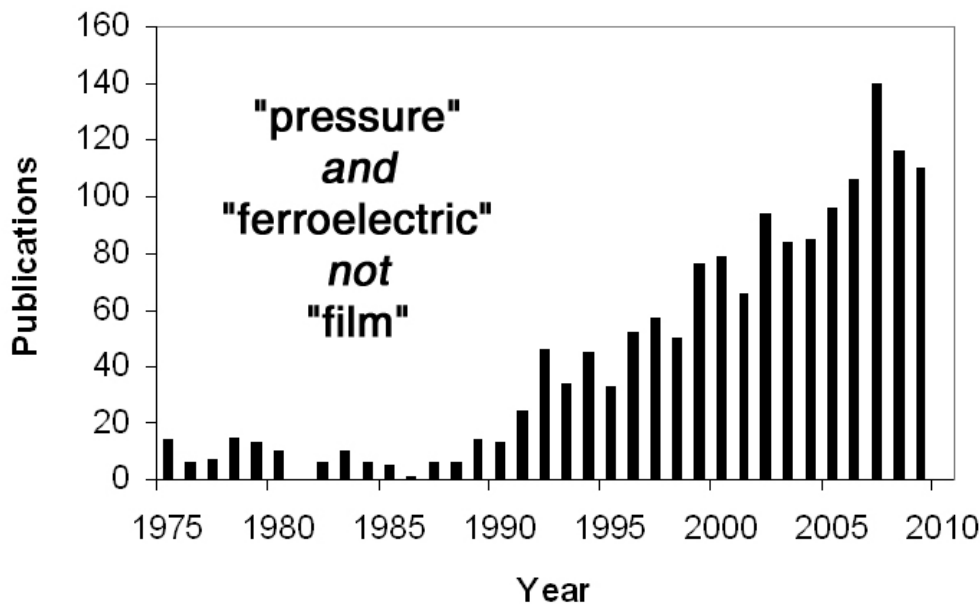

**Figure 3**
Evolution of the number of publications per year containing the keyword "ferroelectric" and "pressure" excluding the keyword "film" to suppress thin film papers with the wording "pressure" in relation to synthesis conditions [37].

Following the pioneering work by Samara *et al.* [163], it was accepted that pressure reduces ferroelectricity in $ABO_3$ perovskites and even annihilates it for a critical pressure $p_c$ at which the crystal structure becomes cubic. Early papers on $BaTiO_3$ [174,175], $KNbO_3$ [176,177], $PbTiO_3$ [178,179] confirmed this view and until recently it seemed unlikely to discover any new fundamental properties and insight from simple perovskites such as the very-well studied $PbTiO_3$. Very recently, however, this perception has changed considerably, as illustrated by the two following studies.

The first striking result comes from ab-initio calculations, published by Z. Wu and R. E. Cohen [180] which indicate theoretically a huge piezoelectric response at high-pressure for a monoclinic *M*-phase in the prototype ferroelectric $PbTiO_3$. The discovery of such a low symmetry *M*-phase is completely unexpected in a pure compound like $PbTiO_3$, and has been found only in the morphotropic phase boundary thus at a specific chemical composition of complex solid-solutions such as PZT, PMN-PT etc. (see section 2). Wu and Cohen's results offer in principle a new avenue by suggesting that $PbTiO_3$ and similar systems under pressure should be the simplest possible system to get to the essential of the underlying phenomena [180]. The very recent literature discusses controversially arguments for [180,181] and against [182-185] such a monoclinic morphotropic phase boundary and the polarization rotation model in pure PbTiO$_3$ (PTO). Future work is clearly needed to clarify this open issue.

A second breakthrough comes through a collaboration that combined progress on both the experimental and the theoretical side, again on PbTiO$_3$ [184,186]. On the experimental side, significant pressure-induced changes in both Raman and X-ray scattering provide evidence for several unexpected structural phase transitions in PTO [184]. The most striking result is that PTO first reduces its ferroelectricity and becomes cubic but then, unlike commonly thought for 30 years [163], becomes again non-cubic through tetragonal-like phases. A deeper understanding of these unexpected result came from the detailed experimental analysis and theoretical ab-initio calculations [184,186]. The study shows that PTO accommodates pressure through mechanisms involving oxygen octahedra tilting and re-entrance of ferroelectricity with the ultimate theoretical prediction that perovskites and related materials enhance their ferroelectricity as hydrostatic pressure increases above a critical value. This new and unexpected high-pressure-ferroelectricity is different in nature from conventional ferroelectricity because it is driven by an original electronic effect rather by ionically-driven long-range interactions [187]. Calculations have namely shown that the underlying mechanism of high-pressure-ferroelectricity in PbTiO$_3$ is the tendency to reduce a high-pressure induced band overlap. According to ab-initio simulations, this pressure-induced phenomenon should occur in *any* insulating transition-metal perovskite as well as other structures involving a transition metal surrounded by oxygen [188]. The experimental verification of the latter prediction will be a main challenge in future high-pressure work on ferroelectrics.

Apart from these two illustrative examples, the effect of high-pressure on dielectric perovskites has attracted an important interest in the past years with a significant amount of experimental and theoretical investigations. Although we do not aim discussing the entire field and its advancements, the interested reader might consult the following (non-exhaustive list of further topics and references of current and future interest.

- Recent X-ray scattering and absorption studies on the effect of high pressure on the *local* structure of ferroelectric perovskites [189-193] have been stimulating. To date such studies remain limited to the study of the well-known ferroelectric - paraelectric phase transition (e.g. p$_c$ = 12 GPa for PbTiO$_3$). These studies should be extended into the very-high-pressure regime (say above 20 or 30 GPa) which, we hope, will provide a better understanding of the recently experimentally observed [184,187] and theoretically predicted [187,188,194] unexpected features.

- There is still a great lack of comprehensive P-T and or/ x-P phase diagrams in perovskite-type solid solutions and this although their complexity can provide a significant understanding of competing interactions. Notable exceptions concern the thorough investigation of the P-T phase diagram of BaTiO$_3$ [195,196] or KNbO$_3$ [177] and the exploration of the x-P phase diagram of the piezoelectric PbZr$_{1-x}$Ti$_x$O$_3$ (PZT), which is characterized by a competition between tilt- and cation-displacements-instabilities [185,197-202]. Such kind of studies should be continued on other materials and the complex phase diagrams are expected to be a mine of understanding of competing instabilities.

- High-pressure investigations of relaxors ferroelectrics and their intrinsic nano-structure have largely contributed to the current interest into the effect of high-pressure on ferroelectrics [203-213]. Among these investigations, diffuse scattering studies have been particularly instructive but remained mostly qualitative and have not yet received a due attention in terms of simulations of diffuse scattering which is still restricted to data at ambient conditions [214-220].

- General rules for predicting pressure-induced phase transitions in perovskites due to ferroelastic octahedral tilting are also of current interest. It was originally suggested by Samara [163] that the phase transition temperatures T$_c$ of zone-boundary transitions in perovskites should always increase with pressure: $dT_c/dP > 0$, i.e. the tilt angle increases with increasing pressure. However, experiments on LaAlO$_3$ [221] and later on other perovskites [222-225] reveal that they undergo tilt phase transitions to higher-symmetry phases on increasing pressure, contrary to the general Samara rule. As a follow-up, Samara's rule have been extended by Angel *et al*. to a new general rule [222] which takes into account the compressibility's of the different polyhedra. However, it remains an open question how these new general rules are respected - or not- in presence of polar cation displacements.
- Finally, recent experiments have illustrated the usefulness of high-pressure for the understanding of multiferroic coupling, e.g. a pressure-induced polarization reversal [226] and pressure-induced enhancement of ferroelectricity [227] have been reported for multiferroic $RMn_2O_5$ manganites. Such experiments should be extended to other multiferroics.

## 5. Strain-Engineering in Ferroelectric Oxide Thin Films

A particular interesting parameter for tuning physical properties in ferroelectrics is the strain, which can be exerted in thin films through differences in lattice parameters and thermal-expansion-behaviour between the film and the underlying substrate. Over the past two decades, a significant amount of progress has been achieved in epitaxial growth of thin films and heterostructures of various complex oxides, including ferroelectrics (see e.g. [228-237]). The literature on strain engineering in ferroelectrics is indeed overwhelming and has led in the recent history, for instance, to the discovery of room-temperature ferroelectricity in strained $SrTiO_3$ [238]. The development of the field of ferroelectric films in terms of publications is illustrated in Figure 4.

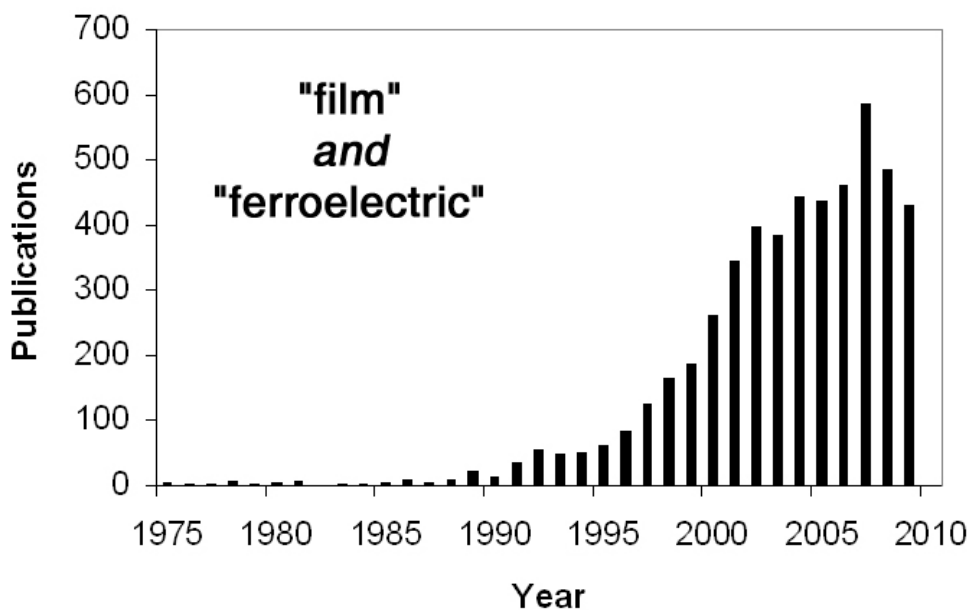

**Figure 4**
Evolution of the number of publications per year containing the keywords "ferroelectric" *and* "film" [37].

     Responsible for this progress are, on the one hand, the maturity of some deposition techniques for the growth of oxides with atomic control, in particular Pulsed Laser Deposition and Molecular Beam Epitaxy [239-241] and, on the other hand, the profusion of useful and realistic theoretical predictions [11,242-251]. These works have pointed out that, due to the large coupling

between polarization and strain, ferroelectrics are among the most interesting materials for strain engineering. By applying epitaxial strain to thin films, the transition temperatures can be increased by hundreds of degrees and new phases can be induced, with symmetries and polarization directions different from the bulk ones. The presence of new phases is interesting, not only because of their potential relevance, but also because they generate new phase boundaries in the temperature-composition-strain space. The presence of temperature-independent phase boundaries (closely parallel to the temperature axis), at which the dielectric and piezoelectric responses are greatly enhanced, is highly desirable [16,65,252].

One of the challenges for attaining this in the lab is the scarce substrate availability. Because the substrate surface structure should be as close as possible to that of the film in the desired orientation, the family of perovskite materials, which offers a large variety of properties and chemical substitutions with a relatively simple structure, is often preferred for strain engineering. High quality single crystals with very particular lattice spacings are required as substrates. These need to be chemical stable at the growth temperatures and should, preferably, be susceptible of presenting atomically flat surfaces. Moreover, not only the lattice parameter at room temperature is important: the thermal expansion coefficients also have to be consistent with the required strain values.

The most widely used substrate for epitaxial growth of complex oxides is (001)-$SrTiO_3$, with a lattice parameter of a= 3.905Å. Some reasons for this popularity are its cubic structure at and above room temperature, its chemical stability and its wide availability. Other available substrates are distorted perovskites such as $NdGaO_3$, $LaAlO_3$, $KNbO_3$, or $YAlO_3$. Fortunately, the choice of substrates has been recently increased by the emergence of rare-earth scandates, from $HoScO_3$ to $LaScO_3$, which offer lattice parameters ranging from 3.93 to 4.05 Å and an excellent crystalline quality [253]. In particular, growth on $DyScO_3$ (a~ 3.945Å) has allowed significant advances in strain engineering, such as the growth of room-temperature ferroelectric $SrTiO_3$ [238], a new phase [254] and periodic domains in $PbTiO_3$ [255] or the control of the domain variants in $BiFeO_3$ [256].

Despite this progress, it is clear that the choice of a particular epitaxial strain presents a serious experimental issue. In order to overcome this limitation, a few groups are working on a more realistic approach that consists of using a combination of substrate and compositional variations. This allows to fully exploit the potential of strain engineering and to fine tune the amount of epitaxial strain to access specific phases or phase boundaries [257-259]. For this method to succeed, *a-priori* theoretical predictions are crucial.

Fortunately, the predictive power of the state-of-the-art theoretical models has proven to be very high [238,251,260,261], and they provide a great aid to design and interpret the experiments. Next to the work of Pertsev and collaborators, which showed the strength of a Landau-Devonshire approach to incorporate strain in a ferroelectric [250], the first principles calculations have also been able to model strain successfully [245,262]. At the moment, a wealth of highly interesting predictions on new materials and new phases are awaiting for experimentalists to be able to test them in the lab [242,257,263].

Other more exotic effects of strain are also of much interest, such as flexoelectricity in the presence of strain gradients [264,265] or the absence of strain-polarization coupling in highly polar ferroelectrics [266].

Magnetic properties are also highly sensitive to epitaxial strain. In particular, in materials with competing ferromagnetic and antiferromagnetic interactions, strain can have drastic effects on the magnetic response [267,268]. Therefore, although relatively little work has been done in this direction, probably with the notable exception of doped-$LaMnO_3$ [269-272] and $BiFeO_3$ [95,267], the effect of epitaxial strain in multiferroic materials is worth to be investigated both from the ferroelectric and the magnetic points of view.

Recently, in different (antiferromagnetic) rare earth manganites under epitaxial strain, with both hexagonal and orthorhombic symmetry, a number of authors have reported splitting between the zero-field-cooled and the field-cooled magnetization vs. temperature curves, which are absent in the bulk material [273-282]. This signals the presence of ferromagnetic interactions in an overall antiferromagnetic film. Spin canting or spin glass behaviour has been proposed but, in most cases, the origin of the induced ferromagnetic interactions is unclear. One of these works, dealing with orthorhombic $YMnO_3$ thin films grown on $SrTiO_3$ substrates, investigates the origin of the ferromagnetic interactions in detail and finds a correlation between the structural distortion and the induced ferromagnetic interactions [275] but if that is the case in all cases is not clear. Other reports find strong evidence that the magnetism may reside at the domain walls [283].

One added difficulty to the experimental investigation of strained films is that, except for some particular cases [284], it involves working with ultra-thin films. In these films, measurements of the structural details or magnetic and electrical responses are very challenging: In the case of magnetic measurements, the substrate magnetic response or minute amounts of paramagnetic impurities in the substrate can mask the response of an otherwise pristine film; in the case of dielectric and ferroelectric properties, leakage is the main obstacle. With respect to the structural characterization, although a lot of information can be obtained using high-resolution diffraction and synchrotron sources, the full structural solution and, in particular, the oxygen atomic positions, of crucial importance to understand the magnetic interactions, cannot be extracted. In ferroelectrics, this limitation is also a crucial one, since the clamping of the substrate can induce internal symmetries that are different from the external lattice symmetry [285]. Moreover, interface [286] and roughness effects [277] also need also to be taken into account when dealing with such thin films.

Either to avoid very large depolarizing fields, in very thin ferroelectric films with out-of-plane polarization, or to relax the epitaxial strain, for relatively larger thickness, the films tend to split into ferroelectric [287] or ferroelastic domains [288], respectively. Moreover, these should appear as periodic patterns, which can be considered as self-assembled ferroelectric bits. In recent years, a few key works have pointed out that the scaling laws that relate the size of the domains with the crystal thickness are valid down to very low thickness, ~50 nm [289,290]. This implies that the size of the periodic domains can be tuned down to a few nanometers by decreasing the film thickness. Comparing domain sizes of ferroelectric and multiferroic materials, insight has been obtained into the energetic of domain wall formation and it has been shown that domain walls in multiferroics, such as $BiFeO_3$ have widths that are slightly larger than those of ferroelectrics (known to be atomically thin) but smaller than those of ferromagnets [291].

Different ferroelectric and piezoelectric responses are expected when ferroelectric domains are that small but more work needs to be done in this direction. Not much is known about how the behaviour of a domain in a patterned film differs from that of an isolated ferroelectric element of the same size [292-294]. Moreover, such small domain sizes entail large domain wall densities. Specially in magnetic and multiferroic films, this implies that a large volume fraction of the film behaves as a domain wall. At present, a few groups are addressing the physical responses of the domain walls, which can show very distinct responses from those of the domains, and the possibility of tuning their density and orientation. In $BiFeO_3$, not only the size of the domains can be changed with the thickness [291] but also deposition on different miscut or type substrates, as well as with different electrical boundary conditions, has allowed controlling the type of domain walls present in the films [295,296]. Conductivity at particular domain walls has been reported for some of these films [105].

In $TbMnO_3$ films deposited on $SrTiO_3$ substrates, despite the large mismatch strain, films are strained up to a thickness of about 60 nm and crystallographic (orthorhombic) domains are observed. In this case, the size of the domains can also be tuned, as it has been found that it increases linearly with thickness [297]. This differs from the quadratic law observed for ferroelectric and ferromagnetic domains, as well as for ferroelastic domains for relatively large thickness, but it is similar to the linear behaviour found in ferroelastic domains for domain sizes of the order of the film thickness [298,299]. Interestingly, in these TbMnO$_3$ films, splitting between the filed-cooled and zero-field cooled magnetization versus temperature curves is also observed and the value of the splitting has been related to the presence of ferromagnetic interaction at the domain walls [300].

Another appealing idea is to deposit a thin film on a substrate for which it is possible to modify the lattice parameter: piezoelectrics. Remind that a piezoelectric material exhibits an electric field (i.e. a voltage difference between opposite sides) through an induced polarization when it is deformed mechanically. The inverse-piezoelectric-effect corresponds to a deformation (thus change in the lattice parameter) when the material is submitted to an applied voltage. This effect is thus in principle exploitable for strain engineering, but is it truly feasible and realistic? The answer is yes, and this has been demonstrated first by K. Dörr's group who has deposited oxides on a single crystalline PMN-PT substrate which exhibits a giant piezoelectric response. After preliminary work of this group starting in 2005 on voltage-controlled epitaxial strain in La$_{0.7}$Sr$_{0.3}$MnO$_3$//Pb(Mg$_{1/3}$Nb$_{2/3}$)O$_3$-PbTiO$_3$(001) films [301,302], a real breakthrough has been their observation of a strain-induced insulator state and giant gauge factor of La$_{0.7}$Sr$_{0.3}$CoO$_3$ films [303]. Considering the enormous potential of this approach it is surprising that only few groups have yet published on the subject, certainly hampered by the yet unsatisfying crystalline quality of the used substrates, but progress is under way [304]. Let us finally point out that PMN-PT substrates are not only piezoelectric but also ferroelectric, which leads to three connected properties: (*i*) a ferroic domain state (*ii*) a ferroelectric hysteresis and (*iii*) electric surface charges. These three features are expected to influence the properties of the deposited films below say 150°C (ferroelectric-paraelectric transition of PMN-PT) and merit a particular attention, a challenge that has not been properly addressed.

## 6. Concluding remarks

The field of ferroelectric materials has in the past played an important role in the field of Phase Transitions and is expected to continue to do so, as judging from the current richness and dynamic of this scientific community. As said above, we did not aspire to cover all fields and references in the fast moving field of ferroelectrics but we hope that our overview of four selected topics transports to the reader some of the current excitement in this very active research area. We remind that there are several hundreds of ferroelectric materials, many of which are not the here discussed oxides, but fluorides, hydrogen bonded ferroelectrics, liquid crystals [305-307], organic/polymer ferroelectrics [308-311], ferroelectric phase-change-materials (PCM's) [312-315] etc., with the last three attracting currently a particular interest. Ferroelectrics: still a bright future ahead.

**Acknowledgements**
The authors would like to thank their past and present collaborators for insightful discussions in the field of ferroelectrics and ferroelectric-related materials.

*comparison with other ferroics*, Physical Review B (Condensed Matter and Materials Physics) **74** (2006) pp. 024115